\documentclass[lettersize,journal]{IEEEtran}
\usepackage{amsmath,amsfonts}
\usepackage{algorithmic}
\usepackage{algorithm}
\usepackage{array}
\usepackage[caption=false,font=normalsize,labelfont=sf,textfont=sf]{subfig}
\usepackage{textcomp}
\usepackage{stfloats}
\usepackage{url}
\usepackage{verbatim}
\usepackage{graphicx}
\usepackage{cite}
\usepackage{amssymb}
\usepackage{amsmath}
\usepackage{url}
\usepackage{mathptmx}                  % use matching math font
\usepackage[svgnames]{xcolor} % change the color of text
\usepackage{soul}
\usepackage{wrapfig} 
\usepackage{hyperref}
\IEEEoverridecommandlockouts
\newcommand{\redout}[1]{}
\newcommand{\rev}[1]{#1}
\newcommand{\taskheader}[1]{\vspace{1mm}\paragraphheader{\textbf{#1}}}
\newcommand{\paragraphheader}[1]{\noindent{\emph{#1}}~}

\newcommand{\mrev}[1]{{#1}}

\begin{document}

\title{LAMDA: Aiding Visual Exploration of Atomic Displacements in Molecular Dynamics Simulations}

\author{\author{Rostyslav Hnatyshyn,
Danny Perez, Gerik Scheuermann, Ross Maciejewski, Baldwin Nsonga
\thanks{R. Hnatyshyn and R. Maciejewski are with Arizona State University, USA. \\E-mail: \{rhnatysh,rmacieje\}@asu.edu.}
\thanks{D. Perez is with Los Alamos National Laboratory, USA. \\E-mail: \{danny\_perez\}@lanl.gov}
\thanks{G. Scheuermann and B. Nsonga are with Leipzig University, Germany. 
	\\E-mail: \{scheuermann, nsonga\}@informatik.uni-leipzig.de.}
}% <-this % stops a space
% according to https://tc.computer.org/vgtc/publications/journal/ we leave this as is
\thanks{Manuscript received xxxxxxxx; revised xxxxxxx.}}


\IEEEpubid{10.1109/TVCG.2026.3652905~\copyright~2026 IEEE}
\maketitle

\begin{abstract}
Contemporary materials science research is heavily conducted \emph{in silico}, involving massive simulations of the atomic-scale evolution of materials.  
Cataloging basic patterns in the atomic displacements is key to understanding and predicting the evolution of physical properties. 
However, the combinatorial complexity of the space of possible transitions coupled with the overwhelming amount of data being produced by high-throughput simulations make such an analysis extremely challenging and time-consuming for domain experts. 
\rev{The development of visual analytics systems that facilitate the exploration of simulation data is an active field of research. While these systems excel in identifying temporal regions of interest, they treat each timestep of a simulation as an independent event without considering the behavior of the atomic displacements between timesteps.}
\redout{To better support such analyses,} \rev{We address this gap by introducing} \textbf{LAMDA}, a visual analytics system that allows domain experts to quickly and systematically explore state-to-state transitions. In LAMDA, transitions are hierarchically categorized, providing a basis for \rev{cataloging} displacement behavior, as well as enabling the analysis of simulations at different resolutions, ranging from very broad qualitative classes of transitions to very narrow definitions of unit processes. LAMDA supports navigating the hierarchy of transitions, enabling scientists to visualize the commonalities between different transitions in each class in terms of invariant features characterizing local atomic environments, and LAMDA simplifies the analysis by capturing user inputs through annotations.   
We evaluate our system through a case study and report on findings from our domain experts.
\end{abstract}


\section{Introduction}
\IEEEPARstart{M}{olecular} dynamics (MD) simulations are a powerful tool in the computational sciences, including materials science, physics, chemistry, and biology. 
\mrev{They provide a fully spatio-temporally resolved view of the nano-scale behavior of materials in terms of the displacements of individual atoms, which otherwise are extremely difficult to observe experimentally. 
The information extracted from these simulations can be used to elucidate basic unit-steps in the evolution of materials, calibrate larger-scale models, or even directly predict the outcome of experiments.}

MD simulations correspond to the solution of a large set of ordinary differential equations (the classical atomistic equations of motion) that describe the evolution of the positions and velocities of atoms in a material. 
The solution of these equations requires an approximation of the interaction energy between atoms, from which the interatomic forces can be obtained and the equations integrated using a simple explicit numerical scheme. 
These simulations generate immense volumes of data, as a single MD trajectory can be thought of as a point evolving in 3$N_\mathrm{atoms}$ dimensions (or even 6$N_\mathrm{atoms}$ when including velocities) over millions of timesteps (a typical timestep being a \emph{femtosecond}, i.e., $10^{-15}$ seconds). This makes the storage, processing, and visualization of MD trajectories a challenging endeavor.

% our particular problem
In many systems of interest, the energy landscape that characterizes atomic interactions contains relatively deep wells that temporarily confine the motion of atoms. 
This makes atomic motion, to a first approximation, a discrete process where long periods of uneventful vibrations are interrupted by infrequent-but-rapid transitions between different wells of the energy landscape. 
The dynamics of a system \mrev{are} further simplified by representing the energy well through a single configuration, often that of the minimum energy configuration within the well~\cite{Wales.2006.PEF}.
This compresses an entire continuous high-dimensional MD trajectory into a much sparser representation in terms of a discrete \emph{state-to-state trajectory} between atomic configurations. 
In spite of this significant simplification, understanding, cataloging, and classifying these transitions is a formidable challenge, given the combinatorial complexity of describing \rev{the collective motion of atoms} in complex systems. 
\rev{This task is extremely important, as changes to the structure of a system can affect its physical and chemical properties. For instance, the shape of metallic nanoparticles directly influences their catalytic properties~\cite{edwards2025size}. Investigating atomic displacements can reveal exactly \emph{how} these changes occur.} 
As such, the development of systematic methods and tools for the exploration and categorization of atomic transitions is a pressing need in the applied computational science community.  

While powerful tools are available to investigate and characterize individual \emph{states} (also referred to as \emph{snapshots})~\cite{hnatyshyn.2023.MS, Duran.2019.VLM, Jurcik.2018.CAA,Skanberg.2018.VVI,Ulbrich.2022.SDM}, the ecosystem of tools available to investigate \rev{the displacement of atoms between states, referred to as \emph{transitions},} is comparatively much less developed. 
Instead, researchers examine transitions one-by-one, using tools such as OVITO \cite{Stukowski.2010.ovito}. 
Based on their findings, analysts then classify these transitions into \mrev{groups}. 
Such a process is prohibitively slow for large datasets, limiting the insights that can be obtained. To our knowledge, there is no tool available to efficiently navigate, compare, and classify ensembles of transitions.
\IEEEpubidadjcol

To address this gap, we present a visual analytics system called \textbf{LAMDA}, a recursive acronym that stands for \textbf{LAMDA Aids Molecular Displacement Analysis}. 
LAMDA organizes transition ensembles through a series of guided data pre-processing steps: a session starts with the interactive reduction of an input dataset, which \mrev{lessens} the cognitive burden on the analyst without \mrev{giving up} valuable information. 
Afterwards, the reduced transition ensemble is organized using a hierarchical clustering approach.
\rev{The results of the clustering can be explored in detail through a suite of multiple coordinated views, ranging from abstract overviews to fully interactive 3D \rev{visualizations}}. 
To further assist exploration, LAMDA supports comparisons between clusters of transitions \rev{through a novel aggregate} 3D representation. \mrev{Insights gained during an analysis can be retained through the system's note-taking features.} 
To evaluate LAMDA's efficacy, we present a case study in which we analyze a large ensemble of transitions alongside our domain expert collaborator. This work contributes the following:  
\begin{itemize}
    \item The application of tensor visualization approaches to encode transitions; 
    \item A view that supports the exploration of hundreds of transitions with fully interactive 3D visualizations coupled with \emph{insight provenance}~\cite{Ragan.2016.CPV} interactions;
    \item A visual analytics system that combines these designs with informative overviews and interactions to support analysts in exploring a transition ensemble.
\end{itemize}
\section{Related Work} 
In this section, we review various methods to analyze long-duration molecular dynamics simulations and discuss the visualization techniques and tools that inspired our system.

\subsection{Quantitative Methods}
\label{sec:related_math}
Atomic configurations of materials are often characterized through the analysis of the spatial relationships between atoms and their neighbors, e.g., the Common Neighbor Analysis~\cite{Tsuzuki.2007.SCD} and the Ackland-Jones Analysis~\cite{Ackland.2006.ALC}. 
These analyses characterize local atomic environments in terms of scalar functions that capture structural characteristics such as local symmetry, enabling the automatic determination of local crystal structure and other basic crystallographic quantities which are commonly used by materials scientists.
Although designed to characterize static structures, these features can also be adapted to characterize transitions by examining how they change across the progression \mrev{from an initial to a final state}, a strategy LAMDA automatically applies to input data \mrev{during the pre-processing stage} (cf. \autoref{data-req}).

The offering of features available to characterize local atomic environments has dramatically increased following the development of modern machine learning techniques to estimate the energy functional used to drive MD simulations. 
Under this approach, the energetic contribution of each atom in the system is approximated as a learnable function of per-atom features that describe the local atomic arrangement in their neighborhood~\cite{Behler.2016.PML, Zubatiuk.2021.DMM, Mortazavi.2023.AMM}. 
These features typically obey the same translation/rotation/permutation invariance as the energy of the system, as the absolute pose of a system in space should not affect the classification of states and transitions.

\subsection{Visualizing Molecular Dynamics Trajectories}
Kocincová et al.~\cite{Kocincova.2017.CVP} introduced a visual analytics system that allows for the comparison of secondary protein structures without the typical occlusion issues presented by 3D visualizations by introducing an abstraction that converts two structures into a sequential representation. 
While this abstraction is powerful and informative, it does not provide information on atomic displacements but rather highlights differences between protein chains; moreover, it is designed for biological molecular systems, which is not the focus of our work.
Mehta et al.~\cite{Mehta.2004.DVA} proposed several techniques that use the 3D locations of atoms as well as the electron density data produced by quantum calculations to evaluate salient isovalues used for isosurface extraction and rendering. 
\rev{Unfortunately, these techniques are limited to analyzing anomalous structures in single atomic configurations and do not consider atomic displacements; moreover, quantum calculations are computationally expensive, especially for long-duration simulations.}
TRAJELIX~\cite{Mezei.2006.TCT} derives scalar values from the differences between a reference helical structure and a MD state and displays these transformations as time-series plots, providing an overview of the transformations that occurred during a simulation. 
\rev{These techniques are limited to the analysis of molecules that have a helical structure (i.e., proteins, ligands, etc.).}
\mrev{VIA-MD~\cite{Skånberg2018VIAMD} highlights regions of interest in bio-chemical molecular dynamics trajectories using a combination of 2D and 3D linked views. Our approach focuses on much smaller systems than the ones explored by VIA-MD which experience major structural changes throughout the trajectory. }
OVITO~\cite{stukowski2009visualization} is widely used among domain experts to visualize single configurations and offers a variety of domain-specific analyses that provide an enormous amount of information\mrev{; it is the tool of choice for our collaborators}.  
However, OVITO visualizations are \mrev{designed to only show one configuration at a time.} 
Analyzing a \emph{transition} entails switching the visualization between its initial and final states without any continuity provided, \mrev{making it a task teeming with potential cognitive overload.}
To the best of our knowledge, no extant systems are designed to extract insights from the temporal relations of atomic configurations. 

\subsection{Visual Cluster Analytics}
LAMDA \mrev{takes advantage of hierarchical clustering algorithms} to form groups of related transitions for the analyst to explore. 
In this section, we review various visual analytics systems that facilitate cluster exploration. 

DICON~\cite{Cao.2011.DIV} generates icons that act as overviews for clusters of high-dimensional data based on the values of its individual members. 
These overviews not only present the cluster's content at a glance but also help viewers evaluate the quality of a cluster. 
Our overview of transition clusters, the \emph{Group Displacement} visualization, is partially inspired by this approach, as we discovered during our design process that experts \mrev{find cluster quality information highly valuable during an analysis.}
VICTOR~\cite{Karatzas.2021.VVAa} is a hierarchical clustering visual analytics system that facilitates cluster comparison through the use of various statistical graphics. 
VICTOR supports variety of different visualizations that provide alternative perspectives on the results of a clustering, and the paper as a whole serves as a valuable design study for developing cluster analytics applications. 
ClusterSculptor~\cite{Nam.2007.CVA} is a general-purpose cluster editing system with coordinated views that support both the analysis and modification of the results of a clustering algorithm. 
Our decision to link the hierarchical heatmap with the dendrograms was inspired by their approach. 
\mrev{Thygesen et al.~\cite{Thygesen.2022.LDE} explored a similar scheme for exploring photochemical transition ensembles using the results of a hierarchical clustering algorithm. 
We note that this paper presents a pipeline and not a fully integrated visual analytics system. 
Besides this, we focus on the \emph{structural} changes between nanoparticle states, while this approach focuses on the \emph{electronic} changes in states.}

% ~\cite{Belghit.2024.CDC} - good survey on VA for MD
\section{Analytical Tasks and Requirements}
LAMDA builds upon our previous work, MolSieve~\cite{hnatyshyn.2023.MS}, a visual analytics system that helps analysts \mrev{discover} temporal regions in MD simulations containing significant structural changes, referred to as \emph{transition regions}. %which contain a large number of individual transitions. 

We discovered that while MolSieve was useful for identifying transition regions, examining and classifying state-to-state transitions was difficult, as MolSieve is designed with single-state analysis in mind.  
Our collaborators would export regions of interest and then manually examine them as ensembles of transitions with external tools. 
While this task was not nearly as laborious as combing through entire simulations, transition regions can still contain thousands of unique and highly complex transitions, posing a significant barrier to their interpretation. 
In this section, we discuss our motivations for building LAMDA in terms of tasks and requirements.

\subsection{Analytical Tasks}
To address this issue, we met bi-weekly for two years to develop LAMDA. 
We adopted an iterative design process during development, working closely with our collaborator, a computational materials scientist with over twenty years of experience. 
Throughout the design process, we identified a set of analytical tasks that reflect the daily workflow of an expert exploring transition ensembles. 

\taskheader{T1: Identify broad \mrev{groups} of transitions.} 
Since molecular dynamics trajectories can be composed of tens of thousands of transitions \cite{perez2016long,perez2018long,huang2018direct}, our collaborators are interested in categorizing transitions into \rev{broad and interpretable classes}. 
Having a robust \rev{categorization} of a system's behavior will enable researchers to predict how it will perform under various conditions, provide a means \mrev{for comparison}, and offer an efficient approach to develop simplified models.

To accomplish this, our collaborators are taking advantage of recent advances in machine learning to cluster transitions. 
Unfortunately, a set of mathematical descriptors (i.e., features) that can accurately partition ensembles of transitions has not yet been identified. 
Thus, our collaborators spend a significant portion of their time writing code to generate clusters of transitions and then examining them using classic visualizations such as heatmaps and dendrograms. 

\taskheader{T2: Identify what characterizes \mrev{groups} of transitions.}
If an analyst decides to examine a potential clustering further based on the information they gathered from high-level visualizations, they resort to using 3D renders to qualitatively describe transitions one-by-one before organizing them into subjective groups based on shared physical characteristics.
This tends to be a tedious and error-prone process due to the cognitive load presented by reasoning across hundreds of transitions, further compounded by the fact that each transition can consist of tens to hundreds of atoms being displaced, \rev{with each transition potentially being oriented on different axes}.

\taskheader{T3: Evaluate clustering quality.}
Analysts need to verify the robustness of the overall clustering, ensuring that it captures the underlying behavior of the system.
Verifying the robustness of a clustering method for transitions is difficult because descriptive statistics do not capture the nuances of atomic displacements, while manual comparison is again impractical due to its overwhelming cognitive burden. 

\subsection{Requirements}
Based on these analytical tasks and their challenges, we derived a set of requirements for a visual analytics system that would streamline the process of generating and examining a new clustering of transitions. 

\taskheader{R1: Generate and interactively explore \mrev{groups} of transitions.} 
\rev{Analysts should be able to directly provide transition data and a set of features as input; the features should reflect some physical component of each transition as they will be used for clustering.}
The clustering should be straightforward to explore yet allow for detailed exploration on demand (i.e., down to individual transitions) \rev{and should guide analysts towards particularly interesting \mrev{groups}}.
\rev{Visualizations of individual transitions should be able to be customized and rendered with external data} for analysis.

\taskheader{R2: Provide an overview for a given \mrev{group of transitions}.} 
Analysts should have a concise overview available for any transition \mrev{group} of interest. 
This overview should not obscure or abstract spatial information and present the salient theme of the \mrev{group} being explored. 
It should also simplify the process of comparing \mrev{groups} of transitions.

\taskheader{R3: Provide a means to evaluate cluster quality.}
Since \mrev{groups} of transitions are identified via machine learning methods, they may include a number of dissimilar transitions.
This could indicate that the clustering should be adjusted, so the system should provide a means to quickly identify when clusters are not uniform. 

\taskheader{R4: Provide an organized way to store insights gained during exploration.} 
Analysts should be able to store their insights, similar to an ordinary notebook. 
The notebook-like approach should integrate with the system and guide experts back towards where they discovered their objects of interest.

\taskheader{R5: Export results for further analysis.}
Analysts should be able to export the insights they gained during the analysis, along with any arbitrary set of transitions. 
All exported data should be \mrev{stored in a} portable format to \mrev{facilitate further exploration and scientific discourse.}

\taskheader{R6: Visual and computational scalability.}
The system should reduce the cognitive load on the expert while remaining highly responsive, even when faced with large \mrev{dataset}s (several thousand transitions).
To support this, redundant transitions should be removed automatically, with the ability to adjust the level of reduction. 

\section{Background}
In this section, we discuss data transformation techniques from the literature that inform our approach.
\rev{\mrev{To begin}, we define an atomic state (i.e., configuration; snapshot) with $n$ atoms as an $n\times3$ matrix $S$, with the $i^{th}$ row corresponding to the 3D coordinates of the $i^{th}$ atom. 
As such, we can view a molecular dynamics simulation as a temporal sequence of $n\times3$ matrices $M = (S_{t},\dots,S_L)$; \mrev{$L$ being the length of the sequence and $t$ the timestep of a state.}}
\rev{A transition $T$ between two states \mrev{$S_0$} and \mrev{$S_1$} is then defined as a tuple $T = (S_0,S_1)$. The set of transitions described by a simulation is $T_{all}=\{T_{t},\dots, T_{L-1}\}$ where \mrev{$T_t$ is a tuple composed of temporally adjacent states $S_t$ and $S_{t+1}$ from $M$}. The $i^{th}$ row in both matrices must correspond to the same atom; i.e., the states must be consistently labeled}. \rev{Finally, a \emph{transition ensemble} is any subset of $T_{all}$.}

% Note that the atom labels are not consistent throughout the set of all transitions.

\subsection{Aligning Atomic Displacements}
\label{sec:alignment}
\rev{As mentioned previously, we are concerned with the analysis of transitions in a MD simulation, i.e., the motion of the atoms from one state to another. 
Visually comparing multiple transitions is difficult because they can be oriented arbitrarily. 
%For instance, given two transitions ($A \to B$, $C \to D$), a similar movement can be observed between two ends of the system; this can be overlooked due to the effects of occlusion and perspective.
% This can occur even if their states share a similar structure and the atoms are labeled consistently between the transitions.
} 
To address this, LAMDA uses alignment techniques to orient transitions as consistently as possible before visualizing them.
\rev{To the best of our knowledge, the alignment of transitions between atomic configurations has not been explored.}

\begin{figure}
  \centering 
  \includegraphics[width=0.8\columnwidth, 
  ]{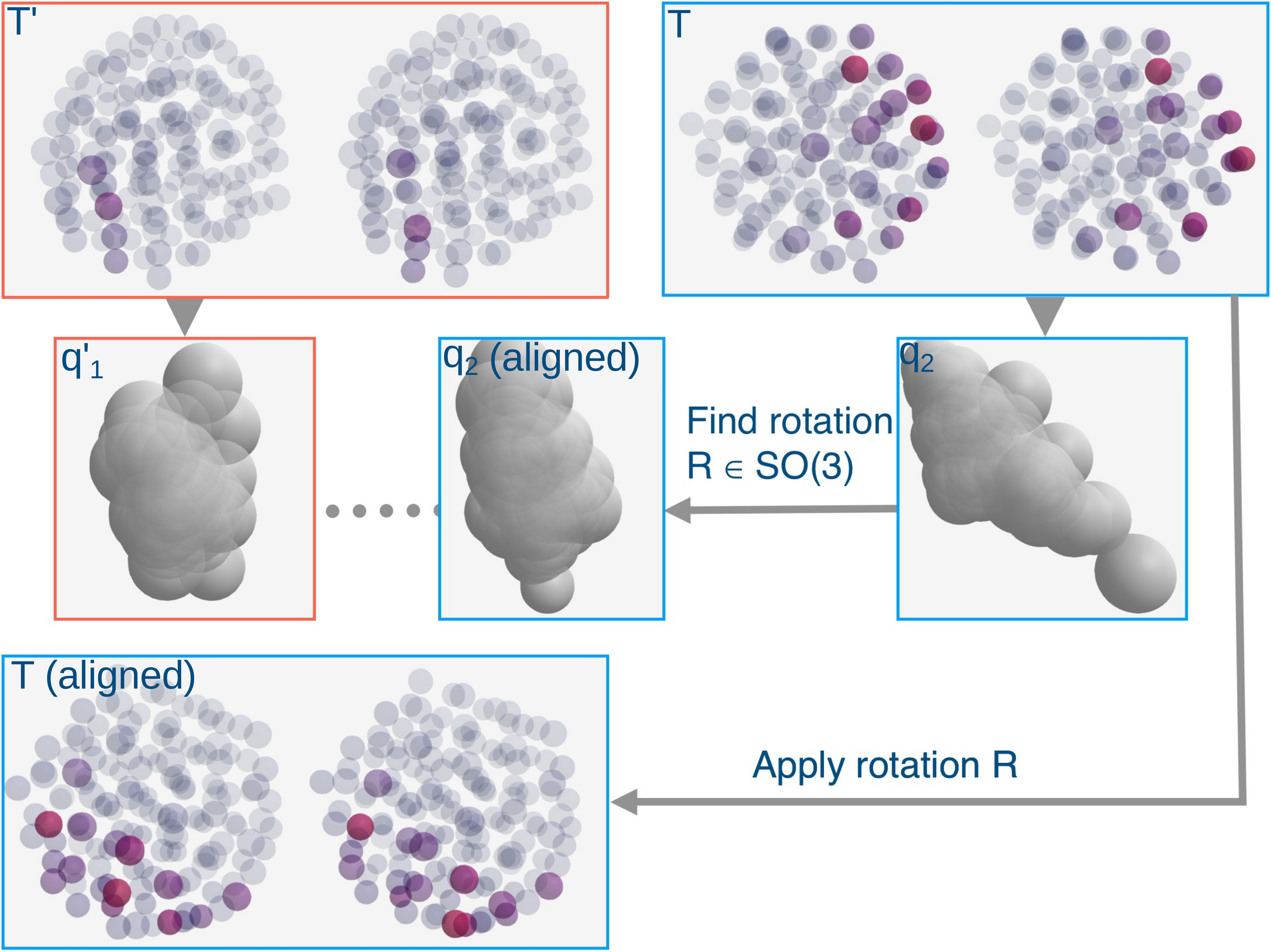}
  \caption{\mrev{Our alignment scheme: the transition \mrev{$T'$} (blue) is being aligned to \mrev{$T$} (orange). 
  Both transitions contain an initial and final state, where atoms are colored by their changes in bond length.}
  These changes indicate the areas that should be aligned. 
  Pseudo positions ($q_i$) are calculated based on \mrev{a transition's $\Delta f$ matrix}. 
  A point set alignment algorithm is then used to calculate a rotation matrix $R$ that results in an appropriate correspondence indicated by the dotted lines. 
  The matrix $R$ is then applied to \mrev{$T'$}. 
  }
  \label{fig:alignment}
\end{figure}

LAMDA's alignment approach (\autoref{fig:alignment}) is composed of simultaneously computing pose and correspondence registrations, as the optimal mapping of the motions in two different transitions is {\em a priori} unknown.
To achieve this, \rev{we first compute a matrix of $n \times k$ per-atom features $f$ for each \emph{state}, where the $i^{th}$ row corresponds to the feature vector for atom $i$}. 
Then, we calculate \mrev{$\Delta f = f_{s_1} - f_{s_2}$}, \rev{the difference between the feature matrices of the initial and final states of transition $T$. 
Note that these features do not necessarily have to be real physical quantities of \rev{an atomistic system}; they \mrev{only need to capture the local structure surrounding the atoms.} 
\mrev{Consequently, $\Delta f$ is non-zero only for atoms whose local environments are affected by the transition.}
Natural choices for $f$ are per-atom features used to train machine-learning-based interatomic potentials, \mrev{as these are specifically designed to capture physically relevant atomic distortions.} 
For instance, we used the Spectral Neighbor Analysis Potentials}~\cite{Rohskopf.2023.fitsnap} as $f$ \rev{for our case study (c.f.~\autoref{sec:case_study}) because they characterize the local atomic environment and are invariant to rotations, translations and permutations.}  

\mrev{With the features prepared, we define two sets of \rev{$k$} \emph{pseudo-atoms} \rev{for each transition, each set corresponding to one state.
The pseudo-atom positions $q_j\ (1 \le j \le k)$ are permutation-invariant versions of the atomic positions $S$.
These can be defined in terms of pseudo-center-of-mass position with $\Delta f$ playing the role of effective masses. Formally, 
\begin{equation}
    q_{j,\alpha} = \frac{\sum^{n}_{i=1} S_{i,\alpha} \Delta f_{i,j} } {\sum^{n}_{i=1} \Delta f_{i,j} }
\end{equation}
where $\alpha$ indexes cartesian directions. \mrev{These values can be interpreted as a normalized sum of atom positions, weighted by the rate of change -- this approximates a ``center of change" for each feature.}
The positions are independent of the ordering of the atoms within each transition, which allows us to focus on aligning atoms that \emph{change similarly} rather than relying on their labels for information.} 
The goal of the alignment is to find a pose which aligns these pseudo-atoms so that the displacement patterns of different transitions can easily be compared.}

\mrev{These pseudo-atoms are then centered so their average position is located at the origin, i.e., 
\begin{equation}
    \hat{q}_{j,\alpha} = q_{j,\alpha} - \frac{1}{n} \sum^{n}_{i=1} q_{j,\alpha} 
\end{equation}}
\mrev{To calculate a possible alignment between two transitions $T'$ and $T$, we use Kabsch's algorithm~\cite{lawrence2019purely} to rotationally align their corresponding $\mathbf{\hat{q}}$'s. 
If a comparison of transitions that is invariant to the exchange of the initial and final states is preferred, the alignment procedure presented above can be carried out a second time after exchanging initial and final states in $T'$. The ordering that minimizes the residual of the alignment is then selected.  We apply this process to align any group $G$ of transitions by arbitrarily selecting a reference transition $T_{ref}$ and aligning all remaining $G$'s members to $T_{ref}$.}
%\rev{To calculate an alignment between two transitions $T'$ to $T$, we select one of the sets of pseudo-atoms for each transition and compute their centers of mass, $C'$ and $C$; each $\mathbf{q}$ is translated to the origin by its corresponding center of mass to ensure the final rotation is applied properly. 
%Finally, we use Kabsch's algorithm~\cite{lawrence2019purely} to obtain a rotation matrix $R$ \mrev{from the two sets of $\mathbf{q}$}. 
%The position matrices in $T'$ are then multiplied by $R$ to apply the rotation. 
%\mrev{We repeat this process with the other set of $\mathbf{q}$ in $T'$ and compare the resulting residuals to yield the optimal alignment between $T$ and $T'$.}}
%We apply this process to align any \mrev{set} $G$ of transitions by selecting a reference transition $T_{ref}$ and aligning all $G$'s members to $T_{ref}$.

\begin{figure}
  \centering 
  \includegraphics[width=0.7\columnwidth, 
  ]{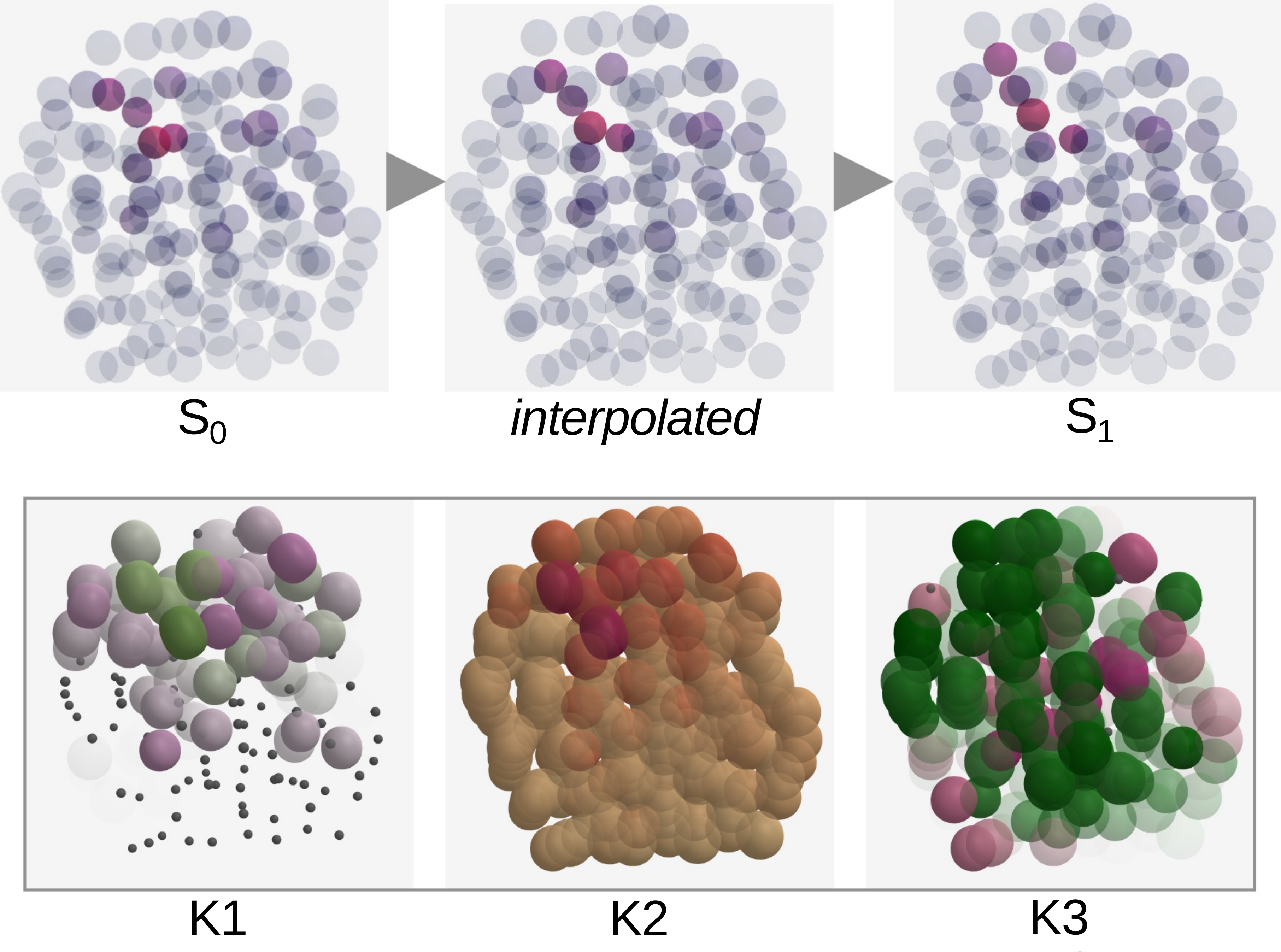}
  \caption{%
  	\mrev{An example of the superquadric visualization; it displays the local displacement around each atom without the need for animations.} The top row of figures illustrates a typical atomic visualization of a transition, while the bottom row \mrev{is the same transition visualized} as a superquadric, \mrev{colored with different strain invariants (\autoref{invariants})}.
    \mrev{$K_1$ is the default value used for coloring superquadrics; the others are included for illustrative purposes as they contain higher-order information about the deformation.}
   }
  \label{fig:superquadrics}
\end{figure}

\subsection{Finite Strain Theory}
\label{invariants}
\begin{figure*}
    \centering    \includegraphics[width=\textwidth]{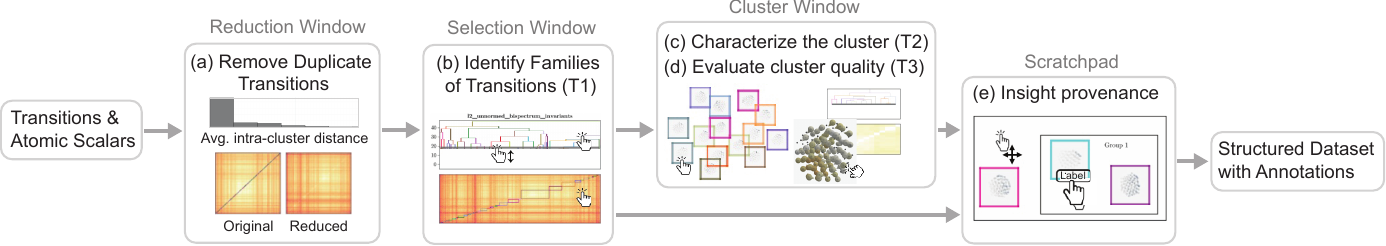}
    \caption{An overview of LAMDA's workflow. Initially, an ensemble of transitions and scalars of interest are provided as input. a.) Analysts interactively remove duplicate transitions in the ensemble and cluster them using the \emph{Reduction Window}. b.) Analysts examine the results using the dendrogram and the heatmap to identify broad \mrev{groups} of transitions (\textbf{T1}). The \emph{Selection Window} supports in-depth exploration of both clusters and transitions through clicking on the dendrogram and heatmap, respectively. (c. \& d.) Clicking the dendrogram displays a \emph{Cluster Window} that provides views and interactions that characterize the cluster (\textbf{T2}) and help evaluate its quality (\textbf{T3}). e.) 
    Transitions and clusters from elsewhere in LAMDA can be stored in the \emph{Scratchpad}, a centralized location for examining and organizing the analyst's selections to generate insights that can be later exported. }
    \label{fig:workflow}
\end{figure*}

% show transform from discrete space into continuous space?
We also investigated visualization schemes that encode the relative displacement occurring during a transition from a continuum mechanical perspective using a method introduced by Gullett et al.~\cite{Gullett.2008.strain_tensor}.
\rev{This consists of computing the Lagrangian strain tensor $E \in \mathbb{R}^{3 \times 3}$ for all atoms in a transition, where $E_i$ encodes the deformation of the neighborhood of atom $i$ without rotation with reference to its initial state.}
We associate the resulting tensors with the initial positions of the transition and calculate scalar values that remain invariant~\cite{Kindlmann.2004.superquadrics} under rigid rotations, referred to as $K_{1,2,3}$:
\begin{equation}
    \begin{split}
        & K_1= \mbox{tr}(E) =\lambda_1+\lambda_2+\lambda_3, \\
        & K_2 = \mbox{norm}(\tilde{E})=\sqrt{\lambda_1^2+\lambda_2^2+\lambda_3^2}, \\
        & K_3=  \mbox{mode}(\tilde{E}) =3\sqrt6 \frac{\lambda_1\lambda_2\lambda_3}{(\lambda_1^2+\lambda_2^2+\lambda_3^2)^{\frac{3}{2}}},\\
    \end{split}
\end{equation}
where $\tilde{E}$ denotes the deviator of $E$ and $\lambda_1$, $\lambda_2$, and $\lambda_3$ denote the eigenvalues of $E$. 
While there are a number of different invariants for three-dimensional second-order tensors\cite{Spencer.2004.continuum_mechanics}, we found these particular invariants to be well-suited for visualization, as they encode easily interpretable physical properties of transitions.

These values are used to generate superquadric~\cite{Kindlmann.2004.superquadrics, Paschalidou.2019.SRL} visualizations of atomic states (\autoref{fig:superquadrics}).
\rev{$K_1$ encodes the volume change (dilation) in the neighborhood of the atom. 
$K_2$ encodes the magnitude of the distortion in an atom's neighborhood; when $K_2 = 0$ (i.e. no distortion), atoms are rendered as regular spheres, indicating isotropic behavior. Since $K_2$ simply quantifies the amount of distortion, we look to $K_3$ to get more information on the nature of the distortion when $K_2 > 0$.}
$K_3$ is the so-called mode of distortion satisfying $-1 \leq K_3 \leq 1$.
\rev{The spheres continuously morph into a rod-shaped glyph for $K_3 = -1$ (linear anisotropy) or a disk-shaped glyph for $K_3 = +1$ (planar anisotropy). All three of these invariants are used to calculate the shape of each glyph \mrev{(\autoref{transition-visualizations}.b)}, while only $K_1$ is used for coloring them.}

\section{LAMDA}
\label{lamda}
LAMDA's workflow (\autoref{fig:workflow}) guides experts through an interactive reduction and clustering process \mrev{before letting them explore the dataset.} 
Analysis is supported at any level of detail, which can range from the abstract perspectives provided by heatmaps and dendrograms (\textbf{T1}) to detailed 3D \rev{visualizations} of individual transitions (\textbf{T2}). 
Notably, LAMDA provides a suite of interactive visualizations that act as an overview of atomic behavior within a group and provide visual indicators of the quality of a cluster (\autoref{transition-visualizations}.c; \textbf{T3}). 
\rev{LAMDA also includes a set of help tooltips for views with special interactions to facilitate on-boarding~\cite{Stoiber.2019.VOL} with the system. 
These tooltips include information on hotkeys, possible interactions, as well as the interpretation of certain visualizations, e.g., superquadrics (\autoref{transition-visualizations}.b).}

LAMDA uses interactive visual linking throughout its interface to mitigate issues with cognitive strain caused by context-switching.
All visual representations of both transitions and clusters are linked to all of their other instances throughout the system. 
To further mitigate cognitive strain and reduce \mrev{visual distances for comparisons}, we included a view for \emph{insight provenance} in the \emph{Selection Window} called the \emph{Scratchpad} (\autoref{scratchpad}), where practitioners can freely position embedded 3D \rev{visualizations} of transitions and clusters to track their analytical process. 
LAMDA is implemented using the Julia programming language and the Makie~\cite{Danisch.2021.Makie} graphics framework as a desktop application, with some of the data pre-processing operations (\autoref{data-req}) being powered by the Atomistic Simulation Environment (ASE) library~\cite{HjorthLarsen.2017.ASE}. 
The color scales used for \rev{most} visualizations were selected from the set of perceptually uniform scales introduced by Kovesi~\cite{Kovesi.2015.GCM}; the heatmaps throughout the interface use the $linear\_worb$ scale, while the dendrograms \rev{use the dynamic color scheme introduced by Tennekes and de Jonge~\cite{Tennekes.2014.TCC} (\autoref{sec:selection_window}).}

\subsection{Pre-processing}
\label{data-req}
\rev{Prior to an analysis in LAMDA, the analyst needs to identify a \emph{transition ensemble} they are interested in. 
Our previous work, MolSieve~\cite{hnatyshyn.2023.MS}, provides a workflow to achieve this, but in principle any method can be used as long as they meet LAMDA's input requirements. There are no special interactions between MolSieve and LAMDA -- experts simply export a transition region they are interested in and view it in LAMDA.} 
LAMDA relies on a minimal amount of input data, which not only greatly reduces the amount of pre-processing that needs to be done before an analysis but also provides a great deal of flexibility.
LAMDA expects two Pickle files per \mrev{dataset}: one that contains a Python dictionary of transition labels to a tuple of ASE Atoms objects; \rev{transition labels are tuples of state IDs that correspond to the initial and final states in the transition.} The other file contains a dictionary of \rev{state IDs} to matrices that represent per-atom \rev{features (\autoref{sec:alignment})}.
\mrev{The matrices for each individual state should be of dimension $n$, the number of atoms in each state, by $k$, the number of features.} 
\mrev{Additional scalar values can be imported to color the 3D visualizations throughout the system (\autoref{transition-visualizations}); these values are not used for clustering.}

LAMDA begins pre-processing by computing the bond connectivity and distances between atomic positions for all of the unique states found in the transition ensemble. Afterward, the \rev{feature deltas}, \rev{changes in bond lengths}, and atomic displacements are computed for each transition.
The \rev{feature delta values are subsequently used to compute the distance matrix used for clustering. 
To do this, each transition's \mrev{$n$ by $k$} feature delta ($\Delta f$; \mrev{\autoref{sec:alignment}}) matrix is collapsed into a $k$-dimensional feature vector by computing aggregate measures using the classic Coulombic potential energy function \rev{where the feature deltas can be conceptualized as pseudo-charges} ~\cite{Cohen.1982.PTE} ($O(kn^2)$); this ensures that the features are aggregated in a way that respects the overall structure of the transition \rev{while being invariant to rigid rotations, translations, or permutations of atoms.} 
Afterward, the distance matrix is constructed by taking the $L^2$ norm between all feature vectors, followed by the ZCA-Whitening transformation~\cite{kessy2018optimal}, which de-correlates the result.} 
\rev{This procedure guarantees that the distance between transitions is not sensitive to their absolute pose.}

Finally, LAMDA computes the global range for each scalar value, used when coloring \rev{visualizations} to enforce visual consistency.
Thanks to the power of ASE, LAMDA is highly flexible and can be used to analyze virtually any type of molecular dynamics simulation; however, we designed LAMDA  around \rev{nanoparticles} (i.e., systems with several hundred atoms), since they typically exhibit very complex state-to-state transitions compared to bulk systems, where geometric and energetic constraints typically limit the size and diversity of transitions. 

% After the data is pre-processed, the alignment is initially applied to states within transitions and is then automatically updated based on the clusters selected by the analyst, where transitions within a cluster are aligned to each other which facilitates comparison between groups of transitions that are similar.
% For each alignment, we compute a \emph{reference transition} -- the median transition for a group computed using the group's distance matrix. 

\subsection{Transition Visualizations}
\label{transition-visualizations}
In this section, we introduce the various 3D \rev{visualizations} for transitions found throughout LAMDA.
There are \rev{three} types of 3D \rev{visualizations} supported by LAMDA: the \emph{Atom Visualization}, the \emph{superquadric Visualization}, and the \emph{Group Displacement Visualization}.
The first two \rev{visualizations} provide complementary perspectives on the changes to the system occurring during a transition (\textbf{R1}), while the \emph{Group Displacement Visualization} provides an overview for arbitrary sets of transitions, which LAMDA uses to represent clusters (\textbf{R2}).

\paragraph{Atom Visualization}
This was our initial approach for visualizing transitions (cf. \autoref{fig:alignment},\autoref{fig:superquadrics}), due to its use in popular tools such as OVITO~\cite{Stukowski.2010.ovito}. 
Each atom is simply rendered in three-dimensional space as a sphere; \mrev{a slider allows experts to interactively interpolate between the initial and final positions of a transition.}
Each sphere is colored using the $linear\_wcmr$ scale~\cite{Kovesi.2015.GCM}, \mrev{which is indexed with the currently selected scalar value, set with a drop-down menu (defaults to average bond delta). The alpha channel corresponds to the scalar value, causing low values to become increasingly transparent as they approach the global minimum.} 
Using transparency serves two purposes: it mitigates occlusion issues caused by atoms being surrounded by their neighbors, and it highlights regions of interest within a transition while reducing unnecessary information. Atoms with high delta values experience changes in their atomic neighborhood, regardless of the scalar chosen. 

While the \emph{Atom Visualization} is effective for analyzing single transitions, it requires significant effort to make sense of a group. 
We felt that we could capture displacements in a transition without relying on animations (i.e., the interpolation between initial and final positions of a transition).
We retained this visualization because of its familiarity to domain experts, but these issues led us to exploring \rev{additional} ways to visualize transitions.

\begin{figure}
\centering
    \includegraphics[width=\columnwidth]{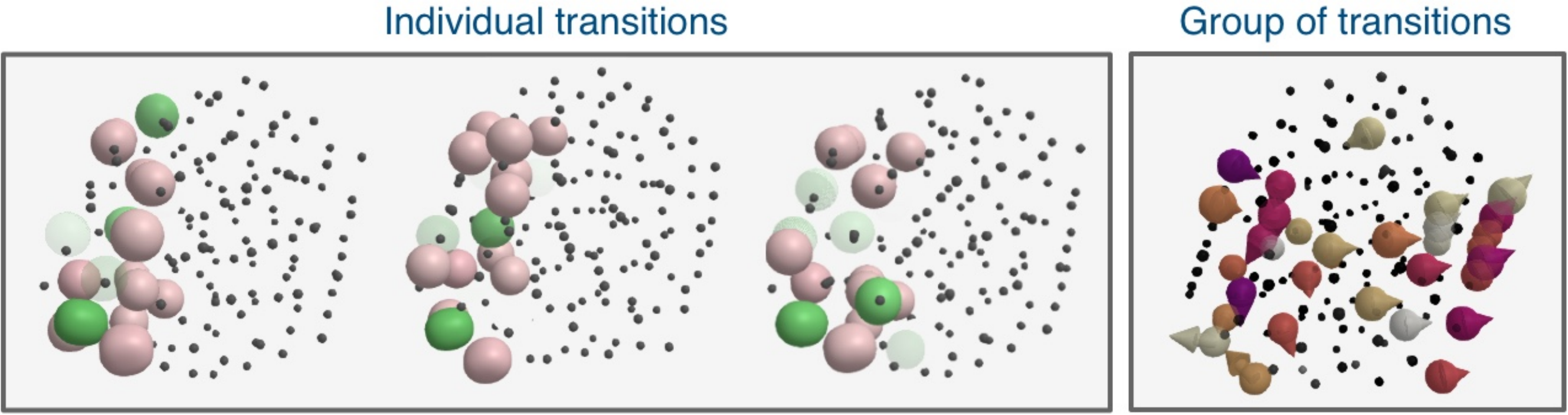}
    \caption{\rev{A visual example of how three distinct transitions are aggregated into the \emph{Group Displacement} visualization. This  visualization is intended to provide overviews for multiple transitions, facilitating intra-cluster comparisons (\textbf{R2}) and providing a visual marker for cluster quality (\textbf{R3}). Saturated colors indicate high displacement correlations between members, while gray spheres indicate the correlation for that point is below an adjustable threshold.}}
    \label{fig:3dvis}
\end{figure}

\paragraph{superquadric Visualization}
\label{sec:superquad}
This approach (\autoref{fig:superquadrics}, bottom; \autoref{fig:3dvis}, \mrev{left}) is designed to demonstrate the transformation of each atom's neighborhood during a transition.
Each atom is rendered as a superquadric~\cite{Kindlmann.2004.superquadrics, Paschalidou.2019.SRL}, which has a variable shape based on the eigen-system of its strain tensor (cf. \mrev{\autoref{invariants}}). 
These eigenvectors represent the principal strain directions. 
By default, $K_1$ is used to color each atom according to the intensity of its extension (green) or contraction (pink) using the $gwv$ diverging color-scale~\cite{Kovesi.2015.GCM}.  
The alpha value of each of element is set according to its intensity, with extreme values at either end of the scale being rendered as increasingly solid. 
Meanwhile, values approaching zero become increasingly transparent, eventually being replaced by small gray spheres to preserve structural information. 
superquadrics inherently encode the directional information of their deformation, thereby eliminating the need for temporal animations to illustrate displacements. 

\paragraph{Group Displacement Visualization}
This visualization (\autoref{fig:3dvis}, \mrev{right}) aims to combine the ease of interpretation provided by the \rev{\emph{Atom} visualization} with the ability to examine multiple transitions at once. 
An attempt to simply plot the constituent atoms of a transition group on top of each other leads to severe clutter, making it difficult to track displacements. 
Instead, we aim to depict the overarching displacement theme of the group based on the coherence between transitions in order to retain spatial information while maintaining visual scalability~\cite{Richer.2022.SV}.
The \rev{\emph{Group Displacement Visualization}} \mrev{highlights atoms that were significantly displaced in a majority of the transitions within a group; the goal is to provide a quick way for analysts to identify major trends in a cluster (\textbf{R2}).} 

\mrev{To determine the atoms that need to be highlighted, the group's \emph{medoid} transition $T_\mu$ is first computed. 
The medoid transition is calculated using intra-group distances, \rev{simultaneously being used as a reference for aligning other transitions.}}

Since atomic positions seldom conform perfectly in practice, even when aligned, we apply a Gaussian kernel to convert the \mrev{initial atomic positions of all transitions within the corresponding group into a continuous space.} 
\mrev{We subsequently sample the displacement for every transition in the group at each position of $S_0$ of the \emph{medoid} $T_\mu$. This corresponds to an interpolation of the displacement using a radial basis function. We then average the results, obtaining $\tilde{\mathbf{d}}$.}

This results in a vector field where displacement regions in phase within the selected group yield high values for $\| \tilde{\mathbf{d}} \|$, while displacements in opposite directions cancel out.
We also calculate a displacement correlation measure~\cite{Vicsek.2010.collective_motion} and rescale it between zero and one, providing a quantitative measure of the overall conformation between displacement regions, readily adapted as a metric to judge the quality of a clustering (\textbf{R3}):
\begin{equation}
      \mbox{corr}(\mathbf{p_i}) = \frac{1}{N}\sum_t \frac{\tilde{\mathbf{d}}(\mathbf{p_i})\cdot \mathbf{d_t}(\mathbf{p_i}) }{ \tilde{\mathbf{d}}(\mathbf{p_i})^2 + \mathbf{d_t}(\mathbf{p_i})^2} + \frac{1}{2},
\end{equation}
where $\mathbf{p_i}$ is a position in $T_\mu$, $\mathbf{d_t}$ is the displacement of transition $\mathbf{t}$ at $\mathbf{p_i}$, and $N$ is the number of transitions in the group.
\mrev{Finally, the initial atoms of the reference transition ($S_0$) are rendered together with arrows that indicate the average displacement direction for that position. Their colors ($linear\_worb$ scale~\cite{Kovesi.2015.GCM}) and alpha values are set corresponding to their correlation values.}
As with the \emph{Atom Visualization}, LAMDA supports interpolating between the initial and final values of the reference transition to provide additional visual context. 

\mrev{Ultimately, our experiments found that the \emph{Group Displacement Visualization} is able to accurately capture the underlying theme of displacement in high-quality clusters.
However, it often conveys less information with larger groups of transitions because the quality of the conformations tend to suffer when vastly different transitions are aligned, causing $\tilde{\mathbf{d}}$ to have low values.
Thus, we note that the \mrev{Group Displacement Visualization} is sensitive to the quality of the input data used to generate the alignment and clusters; it can be used as a canary to indicate that a given clustering may be inadequate.}
To mitigate this issue, LAMDA provides a slider interaction to adjust a threshold value that controls how many atoms and arrows should be rendered; atoms below the threshold are rendered as small gray spheres to indicate their relative lack of movement.

\subsection{Reduction Window}
\label{reduction-window}

\begin{figure}
    \centering
    \includegraphics[width=\columnwidth]{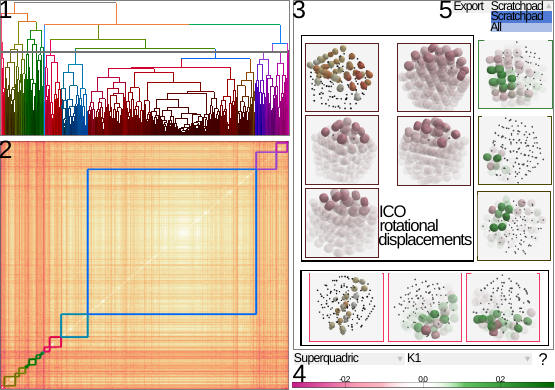}
    \caption{\rev{The \emph{Selection Window}, where experts can explore a transition ensemble organized by cluster. The dendrogram (\textbf{1}) is used to explore the \mrev{dataset}. Experts can click its branches to open a \emph{Cluster Window} to view individual transitions within their hierarchical context. The gray horizontal line is used to update the heatmap (\textbf{2}), which represents the distances between all transitions in the ensemble. The \emph{Scratchpad} (\textbf{3}) allows experts to organize and export their insights through a simple WYSIWYG editor (\textbf{R4}), adding annotations and grouping objects of interest together from selections made across the interface.
    (\textbf{4}) provides interactions that change the visualizations in the \emph{Scratchpad}; (\textbf{5}) shows LAMDA's export options (\textbf{R5}).
    }
    }
    \label{fig:selection_window}
\end{figure}

\mrev{Practitioners begin their analyses with the \emph{Reduction Window}, which is designed to steer the results of the clustering.} 
In this window, experts select a set of feature values to generate a distance matrix (\autoref{data-req}), which is subsequently reduced through an interactive \emph{cleaning by clustering} approach~\cite{Hu.2017.CCM}. 
\rev{As mentioned in ~\autoref{sec:alignment}, the features should reflect the local structure around an atom.
\mrev{A distance of zero between two sets of features indicates that the two transitions share a virtually identical structure.}
Since analysts are interested in only comparing unique transitions, all duplicates save one are removed.} This can potentially improve the quality of the clustering relative to the full \mrev{dataset}; moreover, a smaller \mrev{dataset} translates to an overall lower cognitive burden of analysis as well as increased performance. 

The distance matrix is organized using \rev{agglomerative clustering with ward linkage. The reduction replaces} all of the transitions within a cluster with its \mrev{medoid} (i.e., the transition that has the minimum summed distance to all of its neighbors). \rev{Analysts can steer the reduction by setting a cutoff value for the clustering; groups of transitions with distances below this value are removed. By default, this value is set to one; the reduction can be disabled with a value of zero.}

The \emph{Reduction Window} displays the \rev{full distance matrix as a heatmap on the left-hand side}, neighbored by another heatmap populated with the reduced matrix to the right, with a histogram above them (\autoref{fig:workflow}). 
Both heatmaps are generated by reordering the distance matrices with the Bar-Joseph optimal ordering algorithm~\cite{Bar-joseph.2001.FOL}; \rev{reordering the matrix in this manner places transitions that have low distances to each other nearby in the matrix and matches them to the way the dendrograms throughout the system are ordered.} 
The heatmaps are fully interactive, supporting panning and zooming, which enables detailed inspection. 

To facilitate the search for an optimal cutoff value for the reduction, \rev{the original distance matrix displays groups that will be reduced as rectangles colored from white to black based on the average distances between its members. This causes groups with high distance values to sharply contrast the light colors of the heatmap, alerting the analyst to their presence. Meanwhile, the histogram grounds the color scheme by providing an abstract overview of the intra-cluster distances across all reduction groups. Analysts can use these two visual guides to quickly make a decision about the cutoff value. 
}
When a suitable reduction is found, the analyst can click the \mrev{``Explore"} button to open the \emph{Selection Window}. 
Analysts can return to this screen to readjust the reduction at any time.

\subsection{Selection Window}
\label{sec:selection_window}
\label{selection-window}
The \emph{Selection Window} (\autoref{fig:selection_window}) displays an overview of the reduced \mrev{dataset} with a hierarchical heatmap and dendrogram (\autoref{fig:selection_window}.1-2). 
On the right, the \emph{Scratchpad} (\autoref{fig:selection_window}.3) is shown, which provides an interface for \emph{insight provenance}~\cite{Ragan.2016.CPV} (\textbf{R4}) and export. 

The dendrogram and heatmap provide overviews of the clustering (\textbf{R1}) and support interactions for detailed exploration. 
\rev{As mentioned previously, the dendrogram is colored using a dynamic color scheme~\cite{Tennekes.2014.TCC}. 
This color scheme preserves the hierarchical structure within the coloring by recursively splitting the range of colors that child nodes in the tree can take. 
We initially used a linear categorical coloring scheme because of the huge number of different classes (i.e., each node in the tree would receive a different color). 
While this made it easier to distinguish leaf nodes, it was difficult \mrev{to} trace their ancestry in the hierarchy. The current color scheme preserves the hierarchy and has the added benefit of naturally capturing coarse clusters, but makes it far more difficult to differentiate between neighboring nodes; we discuss how we overcame this problem in \mrev{\autoref{sec:cluster_window}.}}

As in the \emph{Reduction Window}, the heatmap displays rectangles over leaf clusters. 
The cluster cutoff can be adjusted by dragging the gray line in the dendrogram up and down, which automatically updates the rectangles in the heatmap along with the colors of elements in the \emph{Scratchpad} (\autoref{scratchpad}). 
To further aid high-level exploration, the x-axes of the dendrogram and heatmap are linked \rev{since they are ordered in the same way \mrev{(\autoref{reduction-window})}}. 
Panning around in one visualization automatically updates the other to provide a clear link between the dendrogram's abstract representation and the low-level details contained in the heatmap. 
When the dendrogram is hovered, it also displays a tooltip showing the current label of the cluster; this label can be edited \rev{in the \emph{Cluster Window}} (\mrev{\autoref{sec:cluster_window}}).
Most importantly, clicking in either of the two overview visualizations allows analysts to explore transitions and clusters in more detail: clicking in the heatmap embeds the transitions corresponding to the clicked cell in the \emph{Scratchpad} (\autoref{scratchpad}), while clicking branches in the dendrogram opens a \emph{Cluster Window}.

\subsection{Cluster Windows}
\label{sec:cluster_window}
\begin{figure}
    \centering
    \includegraphics[width=\columnwidth]{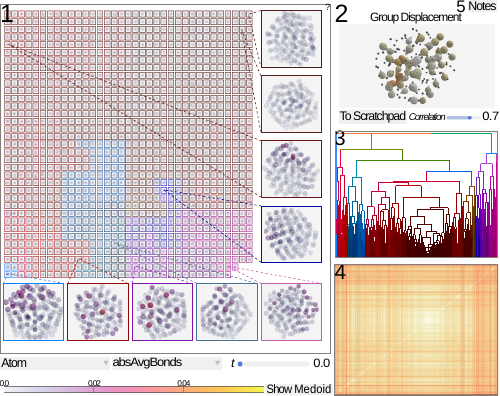}
    \caption{\rev{
    \emph{Cluster Windows} allow for the detailed exploration of transitions automatically clustered by LAMDA (\textbf{R1}).
    The \emph{Embedding View} (\textbf{1}) displays transitions within a cluster as fully interactive 3D visualizations. Transitions are organized in a 2D grid that reflects the overall cluster hierarchy. 
    In this figure, we include a zoomed-out overview of a grid from a large cluster and annotate it with detailed visualizations to provide an illustrative example.
    The \emph{Group Displacement Visualization} (\textbf{2}) provides a 3D overview of a cluster which facilitates analysis and comparisons (\textbf{R2; R3}). 
    The dendrogram (\textbf{3}) provides interactions to navigate the local cluster hierarchy. A heatmap (\textbf{4}) shows distances within the selected cluster; while (\textbf{5}) provides support for annotations and note-taking.}}   \label{fig:cluster_window}
\end{figure}

\emph{Cluster Windows} (\autoref{fig:cluster_window}) are designed for the in-depth analysis of a cluster and its constituent transitions. 
\mrev{Transitions are aligned (\autoref{sec:alignment}) to the cluster's \mrev{medoid} transition prior to rendering to facilitate calculations and visual comparisons}. 
The \emph{Embedding View} on the left (\autoref{fig:cluster_window}.1) provides details about individual transitions (\textbf{R2}), while the views to the right (\autoref{fig:cluster_window}.2-4) display overviews of the cluster's transitions and provide visual cues towards cluster quality (\textbf{R3}).

\rev{LAMDA supports opening multiple \emph{Cluster Windows} to facilitate comparison. 
\mrev{While this is an unconventional approach, expert feedback suggests that being able to interact with only one window at a time would be a significant limitation (\autoref{sec:feedback}). }
This is supported by the work of Plumlee and Ware~\cite{Plumlee.2006.ZMW}, who argue that multiple windows should be used when complex patterns are being compared. 
This is especially relevant for our case, since we are comparing multiple complex patterns at once (i.e., the atomic displacement of hundreds of atoms across dozens of transitions).}

\rev{This window also supports navigating the cluster hierarchy. 
Left clicking branches in the dendrogram (\autoref{fig:cluster_window}.3) switches the window to show that particular node, while pressing $\uparrow$ ascends the hierarchy to show the current cluster's parent. 
Middle clicking a branch opens a new, independent cluster window. 
These interactions are mentioned in the help tooltip for this dendrogram.
Moving the cluster cutoff line adjusts the colors throughout the window, although this may not be useful when examining very small clusters because low-level nodes are colored similarly (\autoref{sec:selection_window}).}
With these general interactions in mind, we can turn our attention to the \emph{Embedding View}.

The \emph{Embedding View} \rev{displays} fully interactive 3D \rev{visualizations} of \rev{the cluster's transitions} (\autoref{transition-visualizations}) in 2D space. 
\rev{We initially tried organizing the visualizations by simply using the UMAP dimensionality reduction algorithm~\cite{Sainburg.2021.PUE} with the cluster's distance matrix as input, as well as using classic multi-dimensional scaling (MDS) algorithms~\cite{Mead.1992.RDM}. 
Besides problems with interpretability posed by UMAP's stochastic nature and the need to tune hyper-parameters, these embeddings suffered from multiple visual issues. 
Transitions would often be rendered on top of each other when they were highly similar -- unfortunately, we realized that modifying the positions of the embedding could potentially affect their interpretation. 
Moreover, the embeddings created by these algorithms would often not match the hierarchical structure of the clustering, making it difficult to judge the quality of a given cluster. 
Finally, scrolling and panning around the embedding view made it difficult to compare individual transitions -- experts would work around this by organizing transitions manually in the scratchpad, often placing them in a grid pattern. 
We realized that an intelligent grid layout could be computed automatically, which led to our current design.}

\rev{Naturally, simply placing transitions in a grid without considering their relationships would only solve the issue of occlusion. Instead, their grid positions are organized according to the cluster hierarchy using a Hilbert space-filling curve (SFC)~\cite{Wattenberg.2005.NSV}. 
The SFC can be viewed as a mapping from 1D to 2D; LAMDA uses the depth-first ordering of the dendrogram's leaves as input to the SFC, which causes neighboring nodes and clusters to be placed near each other. 
Notably, this organizational scheme additionally mitigates the issue of neighboring nodes being colored similarly because their relationships are now encoded spatially.
This is further supported by extensive visual linking, as hovering each transition highlights its position in any dendrogram it appears in (including other windows) and vice-versa.
} 

Inside the \emph{Embedding View}, the analyst can navigate the 2D space to get a closer look at transitions, individually adjust the 3D camera angle within each embedded view, change which \rev{type of visualization} is being used to render the transitions, and adjust what step in the interpolation is being \rev{shown} if supported by the currently selected \rev{visualization} (i.e., the atom and displacement group \rev{visualizations}). 
Changing the \rev{visualization} is supported by a set of buttons, menus, and a slider directly \rev{below} the \emph{Embedding View} (\autoref{fig:cluster_window}.1). 
These adjustments instantly propagate to all visualizations in the window to facilitate comparisons, \mrev{with the exception of individual 3D camera angles, which act independently}. 
Among this set of controls is the \mrev{``Show Medoid"} button which highlights the \mrev{medoid} transition. 
This interaction is intended to provide more insight into the quality of the cluster (\textbf{R3}), as analysts can easily form an expectation of the \mrev{medoid}'s appearance based on the visualizations on the screen and compare it to the rest of the transitions in the group. 

\rev{We believe this view provides a solution to the problems presented by traditional transition analysis workflows because the smallest unit being displayed is now a transition instead of a single state and many transitions can be viewed at once. 
However, the grid layout can still make direct visual comparisons between transitions difficult if many transitions are being displayed. 
As such, this approach works best when multiple grids (i.e., cluster windows) can be displayed, and the analyst focuses on comparing smaller clusters side-by-side.
We also warn that the analytical usefulness of the \emph{Embedding View} is highly dependent on the analyst's input data, much like the clustering and its reduction.
}

Cluster-level annotations can be added by clicking the \mrev{``Notes"} button (\autoref{fig:cluster_window}.5; \textbf{R4}) in the menu bar in the top-right corner. 
This opens another window where analysts can rename the cluster to update the label shown whenever the cluster is interacted with. 
Additionally, a large text area is provided for writing detailed notes; we discuss how these annotations are exported in \autoref{sec:export}.

We ultimately realized there was a need for a separate space to store insights gathered during the analysis.
To this end, we developed the \emph{Scratchpad}, which supports intuitive note-taking interactions, thereby reducing the cognitive burden of comparing data from different clusters located on different screens.
\rev{Double clicking transitions in the \emph{Embedding View} places them in the \emph{Scratchpad}; clusters can also be embedded in the \emph{Scratchpad} by clicking the \mrev{``To Scratchpad"} underneath the \emph{Group Displacement} view (\autoref{fig:cluster_window}.2)}.

\subsection{Note-taking with the Scratchpad}
\label{scratchpad} 
As mentioned previously, the \emph{Scratchpad} (\autoref{fig:selection_window}.3) is meant for organizing insights (\textbf{R4}) that experts can refer to later.
\rev{The \emph{Scratchpad} can be viewed as a simple electronic lab notebook~\cite{Rubacha.2011.REL} directly integrated into the interface. 
While lacking rich interactions, it enables analysts to store information in an intuitive manner. 
We deliberately chose to provide a minimal implementation due to the fact that developing a specific file format to store metadata and other information would quickly become obsolete~\cite{Machina.2013.ELN}. 
Instead, LAMDA can generate a PDF report that captures their insights and directly provides access to the raw data files that analysts are interested in; additional metadata about the simulation can be included directly in the report through the Scratchpad's editing capabilities.} 

As with the \emph{Embedding View}, \rev{3D visualizations are displayed} in a 2D space, with several key differences: \rev{visualizations} are positioned directly by the analyst, clusters can be embedded (\rev{via the \mrev{``To Scratchpad"} button in the cluster window}), annotations can be directly added to the space, and arbitrary groups of transitions can be formed. 
Since the \emph{Scratchpad} does not inherently encode a cluster, transitions are not aligned prior to rendering and the \mrev{``Show Medoid"} button is not available. 

A number of unique interactions in the \emph{Scratchpad} help organize the analyst's thoughts (\textbf{R4}).
Analysts can update the position of a visualization by clicking and dragging the middle mouse button. Double clicking on empty space in the \emph{Scratchpad} creates a textbox where annotations can be entered directly; holding the keyboard button \textbf{T} makes the text bold and is used to denote a title during export (discussed below). 
Analysts can further partition their findings by clicking and dragging in empty space, which draws a 2D rectangle. 
We refer to these as \emph{visual groups}, which are used during the export process. Finally, undesired elements can be removed using the right mouse button. \rev{Again, these interactions are explained in the help tooltip for the \emph{Scratchpad}.}

\subsection{Export}
\label{sec:export}
LAMDA provides two options for export: either by referring to the contents of the \emph{Scratchpad} to generate a custom folder layout or by simply using the clusters indicated by the current height cutoff in the dendrogram. 
Either option can be selected through the export menu in the \emph{Selection Window} (\autoref{fig:selection_window}.5), \mrev{``Scratchpad" or ``All"}.

Both methods of export mimic the hierarchical structure of the data and save transitions as Extended XYZ (.extxyz)%~\footnote{\url{https://open-babel.readthedocs.io/en/latest/FileFormats/Extended_XYZ_cartesian_coordinates_format.html}}
~files, \rev{a common format for molecular dynamics simulations.} 
\rev{These files directly contain the atomic positions of the two states within a transition and can be directly used for further computations.} 
Additionally, a PDF file containing the visual content of the scratchpad \mrev{is created} in the root folder. When \mrev{``All"} is selected, a folder is created recursively for each cluster according to the hierarchy set by the cluster cutoff (\mrev{\autoref{selection-window}}), with transition data being placed in leaf folders. 
If the analyst changed the title of a cluster, its folder is named accordingly, and any annotations created in the \emph{Cluster Window} are saved as plain-text files. 
The \mrev{``Scratchpad"} export option is designed to intuitively map the visual contents of the view to the file system. 
\emph{Visual groups} created by the analyst are exported as subdirectories that support nesting. 
Transition data is placed in its visual group's folder, with transitions outside of visual groups being placed in the root-level folder. 
Text annotations are used either to name the corresponding folder (if in \textbf{bold}, created by pressing \textbf{T} while double clicking) or are placed in a plain-text file.

\section{Case Study}
\label{sec:case_study}
\begin{figure*}
    \centering
\includegraphics[width=\textwidth]{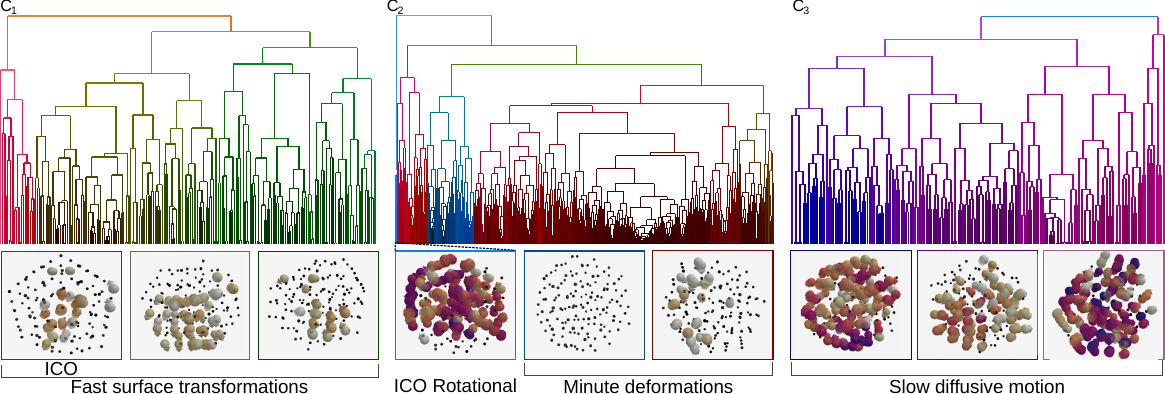}
    \caption{Results from the case study. The dendrograms for each coarse cluster are displayed, labeled with its index and the expert's interpretation; \rev{note that they are not to scale}. At the bottom, representative transitions for each cluster are shown, rendered with \emph{Group Displacement} visualizations. 
    $C_1$ and $C_3$ \rev{are of particular interest, since they contain deformations that significantly alter the structure of the nanoparticle, with these two coarse clusters being separated by their mechanism of action; they either deform the surface or internally re-arrange the structure of the nanoparticle. }  $C_2$ contains most of the transitions of the dataset and is comprised of low-energy transitions \rev{that cause minute deformations on the surface}, partitioned \rev{by magnitude}.}
    
    \label{fig:case_study}
\end{figure*}
% input, discussion about dataset
In this section, we demonstrate the efficacy of LAMDA through a case study conducted as a pair analysis~\cite{arias-hernandez2011pairanalytics} alongside our domain expert (E1) \rev{who is one of the coauthors on this paper}. 
Our case study follows the expert while they examine a \emph{transition region} extracted from a simulation using MolSieve~\cite{hnatyshyn.2023.MS}. 
The original simulation \rev{describes a fixed-size nanoparticle (147 atoms) being subject to a constant temperature (700K) -- in the literature, states from this kind of simulation are often referred to as the NVT (or canonical) ensemble. 
The experiment can be imagined as a nanoparticle floating around in a flask surrounded by a heat bath. 

nanoparticles are often studied due to their useful catalytic properties~\cite{Sharon.2017.PNI}. 
Since nanoparticles are small (at least one dimension is less than 100 nano-meters), they have a high surface area to volume ratio, making their structure susceptible to atomic rearrangements that occur when energy (in this case, heat) is applied to the system. 
\mrev{These rearrangements can lead to significant structural changes which have a direct influence on the nanoparticle's physical properties}. 

\mrev{The transition region we chose to study captures the nanoparticle changing from its initial stable face-centered-cubic (FCC) structure into an icosahedral structure (ICO).}
This structural transformation is well-studied~\cite{huang2018direct, Gao.2016.SOP, Gao.2017.DTP} and is thus a good test-case for both our expert's clustering approach and LAMDA.
While our domain expert already knows \emph{what} is occurring during this part of the simulation, they would like to know \emph{how} it occurs.} Of the eighteen million transitions contained in the original simulation, this \emph{transition region} contains about three thousand. 
While this is a significant reduction, it is still not amenable to manual analysis.

Before they began, the expert computed feature matrices for each transition using the bi-spectrum components used as part of the Spectral Neighbor Analysis Potentials (SNAP) \cite{thompson2015spectral} approach, as computed by the FitSnap~\cite{Rohskopf.2023.fitsnap} package.
\rev{They additionally used the adaptive Common Neighbor Analysis (CNA)~\cite{filot2023bramble} to assign structural labels per-atom in each state and then converted them to per-transition deltas to use as scalar values in the \emph{Atom} visualization (\autoref{transition-visualizations}).  
The analyst believed that these values would help them quickly differentiate between clusters.}
Our expert aimed to investigate the behavior of the system and simultaneously evaluate how well SNAP features cluster, as they inherently encode physical deformations.
% 1.12 provide some numbers on preprocessing the dataset in lamda, how long it took in molsieve 

\subsection{Analysis}
The analyst started by selecting the SNAP features they computed in the \emph{Reduction Window}. 
\rev{Pre-processing took 4 minutes and 6 seconds total, 
10 seconds being used to load the transition data (including scalar values, which are stored as separate files), 33 seconds for computing the transition invariants and 3 minutes and 26 seconds used to calculate the distance matrix.}
Once the distance matrix was computed, they examined the heatmap to gauge if the distances generated would cluster well. 
Noting large groups of distances close to zero, they experimented with different cutoff settings while referring to the heatmap and histogram to ensure the reduction would only remove identical transitions. 
\rev{They settled on \mrev{a} final cutoff value of 0.3, effectively removing nearly half of the \mrev{dataset}, leaving approximately one thousand four hundred transitions (\autoref{fig:selection_window}.2). When asked why they chose this value, they said the histogram in the \emph{Reduction Window} helped them make the decision, as most of the distances were binned at zero.} 

The resulting distance matrix proved consistent with the analyst's expectations due to the relatively long time-span of the transition region in the original simulation, \rev{as well as the fact that transition regions consist of lengthy periods of time where the system transitions back and forth between a number of stable states due to the high amount of energy needed to \mrev{``break out"} of a stable structure~\cite{huang2018direct}.} 
With the reduction complete, they opened the \emph{Selection Window}.

First, the analyst examined the dendrogram and moved its cutoff line to a y-position around 30 to focus on coarse-grained clusters (\autoref{fig:selection_window}.1-2). 
\rev{The clustering initially split the transition ensemble into three distinct clusters: $C_1$ containing red, brown, and green clusters, $C_2$ containing most of the \mrev{dataset}, and $C_3$ containing purple-pink clusters (\autoref{fig:case_study}). 
The analyst inferred that $C_1$ and $C_3$ contained structural changes because they were separated from the rest of the \mrev{dataset}. 
This set of interactions partially fulfills \textbf{R1}, as the analyst was able to generate a clustering quickly and discern which \mrev{groups} of transitions they were interested in exploring using the \emph{Selection Window}.}

\rev{The expert clicked on the red cluster to open a \emph{Cluster Window}. 
The expert immediately recognized that this cluster showed rearrangements occurring to icosahedral structures thanks to the \emph{Group Displacement} view (\autoref{fig:case_study}, left). 
They noticed that the atomic displacements were highly correlated in the center of the system, indicating that the cluster contained highly similar transitions. 
Moreover, the shape of the \emph{Group Displacement} view resembled a nearly perfect icosahedral structure, implying that the transitions were also icosahedral. 
They were able to quickly confirm this by switching the selected scalar value shown in the Atom visualizations to the ICO delta count. 
This displayed a single central atom being highlighted in most of the \emph{Atom} visualizations. 
Some transitions did not have an ICO delta shown but were still grouped together with these transitions; to the analyst, it meant that the clustering was able to capture transitions occurring to structurally similar states that established methods (i.e., the CNA count) fail to recognize (\textbf{R3}). 
This also reflects the difficulty of such an analysis -- numerical methods are still not enough to capture the complexities of atomic displacements. 
The analyst changed the title of the cluster to \mrev{``ICO"} and sent it to the scratchpad. 
Then, they went back to the \emph{Selection Window} to select the red cluster's neighbors, brown and green.}

\rev{Here, the \emph{Group Displacement} view did not have any atoms that displayed a high correlation between transitions; this cluster was noticeably more diverse than the previous one. 
However, they noticed that the heatmap displayed two groups with low intra-cluster distances, one in each cluster. 
When the analyst reviewed the \emph{Group Displacement} visualizations for these groups, they noticed they were highly correlated along one direction on the surface down to the core of the nanoparticle (\autoref{fig:case_study}, left); the analyst interpreted that these groups were both variations on fast surface transformations (FST) (\textbf{R2}). 
These types of transitions dramatically deform the surface of the nanoparticle and re-arrange its structure. 
With this in mind, they switched to the superquadrics visualization for all of the brown-green transitions. 
They noticed that the brown-green clusters differed in magnitude but otherwise were all examples of FSTs.
The analyst attributed their discovery to the superquadrics pinpointing exactly where the atoms were moving the most without needing to track dozens of displacements (\textbf{R1}). 
Due to its distance, the analyst realized the \mrev{``ICO"} cluster also displayed FSTs.
They visually confirmed this via the \emph{Scratchpad} and created a \emph{visual group} around the three clusters called \mrev{``FST"}. }

\rev{Before moving onto the purple-pink cluster on the other side of the dendrogram, the analyst noticed a small light blue cluster on the left side of $C_2$.
They opened a new \emph{Cluster Window} and adjusted the time slider to find they were all rotational displacements occurring to an icosahedral structure. 
These movements are sometimes accompanied by internal re-arrangements within the system's structure, which could be seen in the \emph{Group Displacement} view (\autoref{fig:case_study}, center; \textbf{R2}).
The fact that these were clustered separately lent some level of credibility to the quality of the clustering, as this indicated that transitions weren't simply being partitioned based on the magnitude of their movement alone. 
They sent this cluster to the Scratchpad and annotated it accordingly.}

\rev{The analyst turned their attention to the cluster furthest away from the ones they already investigated, avoiding the large cluster because they believed it would be full of low-energy transitions. 
This is because transition regions often are punctuated by transitions that oscillate between slight deformations of stable structures. 
This led them to open a new \emph{Cluster Window} for the set of purple-pink transitions. 
The transitions here seemed to affect the structure drastically but were mixed in terms of how dramatic the atomic displacements were. 
To understand this, they located a small group within the cluster that was highly similar and reviewed its \emph{Group Displacement} visualization. 
This revealed high correlations in the interior of the system which indicated that these transitions seemed to reconfigure atoms internally (\autoref{fig:case_study}; right).}

\rev{The analyst wanted to compare this group and the FST group from before, so they opened a \emph{Cluster Window} for it. 
Just by moving the time slider back and forth, the difference was apparent -- the FST group deformed the surface, while the new cluster would re-arrange atoms internally. 
The analyst annotated this set accordingly and sent it to the scratchpad.
Finally, the analyst decided to take a look at the rest of the transitions in the ensemble; to do so, they pressed $\uparrow$ to switch the \emph{Cluster Window} to show the entire \mrev{dataset}.
}

\rev{The analyst was impressed by the fact that LAMDA remained responsive even though the entire \mrev{dataset} was rendered at once (\autoref{fig:cluster_window}.1; \textbf{R6}), \mrev{one thousand four hundred fully-interactive 3D visualizations were available for the analyst to explore in real-time.} 
They zoomed in on the purple-pink clusters in the grid and investigated their neighborhood. 
They quickly switched through the various scalar values they created to view the other clusters; as expected, these did not contain transitions with internal structural reconfigurations. 
This was because most transitions did not contain atoms with deltas for either structural label, being rendered as gray spheres.
Switching to the bond delta values did show a number of transitions that had moving atoms, but panning around and looking closely revealed that these were all surface level movements (\autoref{fig:case_study}, center). 
The analyst did notice there were four transitions in the dark red cluster that exhibited rotational behavior because of how much their atomic bond deltas stood out in relation to their neighbors. They sent these transitions to the Scratchpad to examine them side-by-side. 
Ultimately, they found that these transitions did not meaningfully change the internal structure of the system (\autoref{fig:selection_window}.3).}

\rev{Upon reviewing their notes in the scratchpad, the analyst affirmed that the clustering was able to cluster transitions well (\textbf{R4}), separating surface movements from internal displacements by the types of structures they were occurring to. The pink-purple clusters contained transitions that reconfigured interior atoms, while the initial red, brown, green, light blue clusters contained transitions with dramatic surface transformations. Meanwhile, the middle clusters consisted of slight surface level deformations in the crystalline structure partitioned by magnitude. }

This case study demonstrates LAMDA's \rev{exploratory} capabilities. With LAMDA, our expert was able to quickly interpret the results of clustering transitions with SNAP potentials (\textbf{T1} and \textbf{T2}), becoming increasingly convinced of each cluster's quality (\textbf{T3}) throughout the session. 

\subsection{Expert Feedback}
\label{sec:feedback}
\rev{We solicited feedback from the expert (E1) that performed the case study for suggestions on what other elements of the system could be improved. 
Since they were directly involved in the development of the system, we invited another domain expert (E2), a computational materials scientist with over a decade of experience, to simply interact with the system and evaluate its visual components. 
E2 did not have a \mrev{dataset} of their own to explore but was able to explore the data showcased in our case study. 
During both interviews, we discussed how LAMDA compared to their regular workflow and solicited feedback; their comments are interleaved to avoid redundancies.}

\rev{As we mentioned previously, domain experts typically investigate transition ensembles by manually sifting through data extracted by MolSieve with external visualization tools such as OVITO~\cite{Stukowski.2010.ovito}.
Since OVITO renders single atomic configurations, they typically load states in a sequence and flip back and forth between them to investigate atomic displacements. 
As such, both analysts found the \emph{Atom} visualization familiar and intuitive; they appreciated the ability to view multiple transitions at once, as well as the ability to interpolate between the beginning and end of transitions. 
However, E1 found that tracking multiple moving atoms across many transitions was difficult, which is why they turned to the \emph{Superquadric} visualization during the case study. E1 thought that the \emph{Superquadric} view was significantly more abstract, but found it helped them quickly pinpoint transitions of interest within a cluster.
However, they found that this abstraction came at the cost of being able to discern atomic structures, causing them to switch between the visualizations as needed. 
E2 shared these concerns and added that the \emph{Superquadric} visualization required a significant amount of exposition before it could be understood. This suggests that future work could explore alternative abstractions for transitions using the lessons we learned here.
}

\rev{Overall, both analysts were very pleased with LAMDA's performance, both in terms of the system's responsiveness and its streamlined analysis experience.
Both experts found the \emph{Embedding} view valuable because it was able to simultaneously present a large amount of transitions at once and preserve the hierarchy of the clustering. E1 noted that comparing far-away transitions in one \emph{Embedding} view was difficult -- they relied on opening multiple cluster windows to compare them side-by-side.}
\rev{Both experts appreciated the \emph{Scratchpad}, as it provided additional dimensions to their regular notes. E1 said, ``I usually just sit and write things down if I find something, but adding a visual dimension to it really makes things clear, especially because it links right back to the data [when exported].'' E2 said they often take screenshots and annotate them in another program. They thought that being able to interact with the transitions they saved was a significant improvement over static images.}

\rev{One major issue we have discussed throughout this work is the overlapping color scheme for clusters. When we asked E2 about it, they did not find it to be much of an issue -- ``They could be all black frames, the sorted grid solves this issue by separating them by position so I can compare groups easily." E1 echoed this sentiment, but noted that the color scheme could be confusing in the \emph{Scratchpad} where transitions and clusters are removed from their immediate context.
}

When asked about potential improvements we could make, \rev{E1} suggested we could add support for visualizing per-transition scalar values, providing an additional dimension to compare clusters with. 
They provided an example from the case study: if we visualized the changes in potential energy per transition somewhere, the analyst would have easily identified $C_2$ as a set of low-energy transitions. 
Another suggestion was enabling analysts to explore and compare multiple clusterings for one transition ensemble, which could help identify useful features and provide alternate perspectives. 
To add to this, they suggested the clustering itself should be editable through merge and difference operations, which could be used to manually improve cluster quality.
\rev{E1 also mentioned they would like the time slider to be linked across windows, alongside the ability to control the camera for multiple windows in one interface rather than updating cameras individually. They also mentioned having the ability to hide the other plots in the Cluster Window as potentially being useful for side-by-side comparisons.}
Finally, they suggested introducing a plugin system for analysts to experiment with different alignment and distance calculation algorithms, as well as general-purpose programmatic access to individual transitions within the system during run-time.

\section{Conclusion}
In this work, we introduced a visual analytics system called LAMDA that facilitates the exploration and in-depth analysis of a set of transitions. Through its streamlined exploration process and coordinated views, experts are able to quickly \rev{catalog} atomic behavior. The annotation features provide ways to organize their insights, directly linking with LAMDA's powerful export features to help experts quickly iterate on new experimental code based on their interpretations. We verified LAMDA's effectiveness through a case study, \rev{where an expert was able to identify and explore \mrev{groups} of atomic behavior within an ensemble of transitions exported from another tool.} LAMDA is essentially a \mrev{``last mile"} analysis tool for molecular dynamics simulations: while other tools provide coarse descriptions of molecular behavior, LAMDA takes these coarse results and provides detailed insights down to individual transitions. The source code is available at \url{https://github.com/rostyhn/LAMDA}.

% limitation of only a small subset of chemistry? I think the alignment and clustering would still work, but we don't really do any calculations based on special stuff like in organic chemistry
% broad area of work: tracking multiple moving 3D objects across multiple visualizations; general tie in to VIS
We conclude with a discussion of some of LAMDA's limitations that inform potential directions for future work. 
\rev{As we have already discussed, the color scheme still faces some issues with classes overlapping. One solution could be to implement an alternative dynamic color scheme such as the one proposed by Chen et al.~\cite{Chen.2025.DCA}.
Alternatively, we could experiment with} glyphs that provide a unique visual fingerprint for each cluster using information from its transitions.
We also found that the superquadric visualization needs improvement, as small variations in values did not translate to perceptible changes in length, key for comparing shapes~\cite{Qi.2022.QRL}.  
We attempted to compensate for this by exaggerating the distortion of each glyph with constants, but it remains difficult to compare glyphs that slightly vary in value, especially when the values themselves are small. 
\rev{The issue we mentioned with structures being difficult to discern suggests that the perceptual trade-offs between concrete (\emph{Atom}) and abstract (\emph{Superquadric}) visualizations should be investigated in the context of scientific visualization.
In a more general sense, addressing the issue of perceptually tracking multiple 3D objects across many visualizations is a promising and unexplored area of research.}
In terms of versatility, LAMDA was designed with materials molecular dynamics simulations in mind; therefore, our visualizations may not be easily interpretable for molecules that are very large \rev{or contain atoms with different elements}, like the ones found in biological systems (e.g., proteins).
To support the \rev{analysis of larger systems}, specialized techniques that reduce and aggregate atomic displacement information will need to be developed. 
\rev{Supporting atoms with different elements, especially those with special interactions (e.g., forming functional groups) would require an entirely different set of techniques than the ones LAMDA provides.}
Finally, we do not claim that LAMDA is scalable, \rev{as we only tested the system with a little more than a thousand transitions being rendered at once.} With this being said, LAMDA was built on the assumption that experts would be analyzing \rev{\mrev{dataset}s around this size.}

\section*{Acknowledgments}
The authors acknowledge the financial support provided by the Federal Ministry of Research, Technology and Space of Germany and by Sächsische Staatsministerium für Wissenschaft, Kultur und Tourismus in the programme Center of Excellence for AI-research ``Center for Scalable Data Analytics and Artificial Intelligence Dresden/Leipzig“, project identification number: ScaDS.AI

\bibliographystyle{IEEEtran}
\bibliography{bibliography}

\vfill

\end{document}